\documentclass[pra,showpacs,bibnotes]{revtex4}
\usepackage{graphicx}
\begin{document}
\draft
\def\ds{\displaystyle}
\title{Vortex Dipole in a BEC with dipole-dipole interaction}
\author{C. Yuce, Z. Oztas}
\address{ Physics Department, Anadolu University,
 Eskisehir, Turkey}
\email{cyuce@anadolu.edu.tr}
\date{\today}
\pacs{ 03.75.Nt, 03.75.Kk, 67.85.De}
\begin{abstract}
We consider single and multiply charged quantized vortex dipole in
an oblate dipolar Bose Einstein condensate in the Thomas-Fermi
(TF) regime. We calculate the critical velocity for the formation
of a pair of vortices of opposite charge. We find that dipolar
interactions decrease the critical velocity for a vortex dipole
nucleation.
\end{abstract}
\maketitle

\section{Introduction}

A vortex dipole is a pair of vortices of equal and opposite
circulation situated symmetrically about the origin. Under linear
motion of a localized repulsive Gaussian potential, a vortex pair
formation with opposite circulation is possible if the potential
is moved at a velocity above a critical value \cite{jack}. In the
experiments \cite{int13,int40}, a laser beam focused on the center
of the cloud was scanned back and forth along the axial dimension
of the cigar shaped condensate. Vortices were not observed
directly, but the strong heating only above a critical velocity
was measured. It was shown that the measurement of significantly
enhanced heating is due to energy transfer via vortex formation
\cite{63}. Recently, experimental observations of singly and
multiply charged quantized vortex dipoles in a highly oblate BEC
with $\ds ^{87}Rb$ were reported by Neely \textit{et al.}
\cite{ilkdeney5}. In the experiment, vortex dipoles were created
by forcing superfluid around a repulsive Gaussian obstacle using a
focused blue-detuned laser beam. The beam was initially located on
the left of the trap center and the harmonic potential was
translated at a constant velocity until the obstacle ends up on
the right of the trap center. At the same time, the height of the
obstacle is linearly ramped to zero, leading to the generation of
a vortex dipole that is unaffected by the presence of an obstacle
or by heating due to moving the obstacle through the edges of the
BEC where the local speed of sound is small. Vortex dipoles were
observed to survive for many seconds in the condensate without
self-annihilation. The experiment also provided evidence for the
formation of multiply charged vortex dipoles. The authors in
\cite{ilkdeney5} noted that the theoretical predictions of
critical velocity for vortex pair formation in \cite{enerji6} are
in good agreement with the experimental results. The critical
velocity is given by the minimum of the ratio of the energy to the
momentum of the vortex dipole
\begin{equation}\label{mknjkbj}
v_{c}=min(\frac{E(I)}{I})
\end{equation}
where $\ds E(I)$ is the energy of an elementary excitation with
linear momentum (or impulse) $\ds I$ \cite{int5}. When the object
moves at a velocity above a critical value, the superfluid flow
becomes unstable against the formation of quantized vortices,
which give rise to a new dissipative regime
\cite{int24,int41,int14}. Pairs of vortices with opposite
circulation are generated at opposite sides of the object.
Recently, instead of removing the trapping potential and expanding
the condensate to make the vortex cores optically resolvable,
Freilich et al. experimentally observed the real-time dynamics of
vortex dipoles by repeatedly imaging the vortex cores \cite{fre}.
Vortex tripoles have also been observed experimentally
\cite{fre2}. Several theoretical investigations have been reported
for the generation \cite{int15,int16,int17,int18,pekko}, stability
\cite{int19,int20}, and stationary configurations of vortex
dipoles \cite{int21,int22}. In addition, fully analytic
expressions of the angular momentum and energy of a vortex dipole
in a trapped two dimensional BEC were obtained
\cite{int23}.\\
The successful realization of Bose-Einstein condensation of
$\ds{^{52}Cr}$ atoms has stimulated a growing interest in the
study of BEC with nonlocal dipole-dipole interactions
\cite{deney01,deney02,deney03,rev1,rev2,enerji1,enerji2}. A
particular interest is the vortex structures in dipolar
condensates
\cite{int35,int37,int39,int36,int29,int30,cem,kal0,kal,abad0,abad1,int28,int31,int32,int33,int34,int38}.
In contrast to the isotropic character of contact interaction,
long ranged and anisotropic dipole-dipole interaction has
remarkable consequences for the physics of rotating dipolar gases
in TF limit. It was shown that, in axially symmetric traps with
the axis along the dipole orientation, the critical angular
velocity, above which a vortex is energetically favorable, is
decreased due to the dipolar interaction in oblate traps
\cite{int31}. It was discussed that the effect of the
dipole-dipole interaction is the lowered precession velocity of an
off-center straight vortex line in an oblate trap \cite{int32}.
Our aim in the present work is to calculate the critical velocity
for vortex dipole formation in a dipolar oblate BEC in TF regime.
This paper is structured as follows. Section II reviews
Bose-Einstein condensates with dipole-dipole interaction. Section
III investigates the critical velocity in the presence of the
dipole-dipole interaction. The last section discusses the results.

\section{Dipolar BEC}

Consider a BEC of N particles with magnetic dipole moment oriented
in the z direction. In the mean field theory, the order parameter
$\ds \psi (\textbf{r})$ of the condensate is the solution of the
Gross-Pitaevskii equation (GPE) \cite{enerji1, enerji2}
\begin{equation}\label{hjhjg}
\left(- \frac{\hbar^{2}}{2 m} \nabla^{2 }+V_{T}+g |\psi
(\textbf{r})|^{2} + \Phi_{dd}\right) \psi (\textbf{r})= \mu \psi
(\textbf{r})
\end{equation}
where $\ds \mu$ is the chemical potential, $\ds g=\frac{4 \pi
\hbar^{2} a_{s}}{m}$,  $\ds a_{s}$ is the s-wave scattering
length, $\ds V_{T}$ is the trap potential
\begin{equation}\label{ölmýy}
V_{T}=\frac{1}{2}m \omega_{\bot}^2 \left( \rho^{2}+\gamma^{2}
z^{2}\right)
\end{equation}
$\ds \gamma$ is the trap aspect ratio, and $\ds \Phi_{dd}
(\textbf{r})$ is the dipolar interaction
\begin{equation}\label{kknmny}
{\Phi_{dd}(\textbf{r})=\frac{3}{4 \pi}  g \varepsilon_{dd}\int d^3
r' \frac{1-3 \cos^2 \theta}{|\mathbf{r}-\mathbf{r^{\prime}}|^3}
|\psi' (\textbf{r})|^2}
\end{equation}
where $\ds r-r'$ is the  distance between the dipoles, $\ds
\theta$ is the angle between the direction of the dipole moment
and the vector connecting the particles, and the dimensionless
quantity $\ds \varepsilon_{dd}$ is the relative strength of the
dipolar and s-wave interactions \cite{int25}. The BEC is stable as
long as $\ds -0.5<\varepsilon_{dd}<1$ in the TF limit \cite{int25,
int35}. In the regime $\ds \varepsilon_{dd}<0$, the dipolar
interaction is reversed by rapid rotation of the field
aligning the dipoles \cite{int37, int39}.\\
We shall use scaled harmonic oscillator units (h.o.u.) for
simplicity. In this system, the units of length, time, and energy
are $\ds \sqrt{\frac{\hbar}{2 m \omega_{\perp}}}$, $\ds \frac{1}{2
\omega_{\perp}}$ and $\ds \hbar \omega_{\perp}$, respectively.
Hence the GP equation in h.o.u reads
\begin{equation}\label{hjhjg}
\left(-\nabla^{2 }+V^{\prime}_{T}+g^{\prime} |\psi^{\prime}
(\textbf{r})|^{2} +\Phi_{dd}^{\prime}
\right)\psi^{\prime}(\textbf{r}) = \mu^{\prime} \psi^{\prime}
(\textbf{r})
\end{equation}
where the dimensionless interaction parameter $\ds g^{\prime}$ is
given by
\begin{equation}\label{lkknjbn}
g^{\prime}=  \frac{2m N}{\hbar^2 } \sqrt{\frac{2  m
\omega_{\perp}}{\hbar}}~g
\end{equation}
and $\ds{\Phi_{dd}^{\prime}=\Phi_{dd}(g{\rightarrow}g^{\prime})}$.
The normalization of $\ds \psi' (\textbf{r})$ chosen here is $\ds
\int d^{3}r |\psi^{\prime} (\textbf{r})|^{2}=1$.\\
The equation (\ref{hjhjg}) is an integro-differential equation
since it has both integrals and derivatives of an unknown wave
function. The solution of this equation was presented by Eberlein
 \textit{et al.} in TF regime \cite{int25}. They showed
that a parabolic density remains an exact solution for an
harmonically trapped vortex-free dipolar condensate in TF limit:
\begin{equation}\label{nbg}
n(\textbf{r})=n_{0}\left
(1-\frac{\rho^{2}}{{R}^{2}}-\frac{z^{2}}{{L}^{2}}\right)
\end{equation}
where $\ds n_{0}=\frac{\mu^{\prime}}{g^{\prime}}$ and $\ds R$ and
$\ds L$ are the radial and axial sizes of the condensate,
respectively. The condensate aspect ratio $\ds\kappa$ is defined
as $\kappa=\ds{\frac{R}{L}}$. In the absence of dipolar
interaction, the condensate aspect ratio $\kappa$ and the trap
aspect ratio $\gamma$ match. $\ds\kappa$ decreases with increasing
$\ds {\varepsilon_{dd}}$ in an oblate trap. Hence, for $\ds
\varepsilon_{dd}>0$ ($\ds \varepsilon_{dd}<0$), $\ds\kappa<\gamma$
($\ds\kappa>\gamma$) \cite{int35, int37}. In h.o.u, the dipolar
mean-field potential inside the condensate in the TF approximation
is given by \cite{int25}
\begin{equation}\label{joýhn}
\Phi_{dd}(\textbf{r})=n_0g^{\prime}\varepsilon_{dd} \left(\frac{
\rho^{2}}{R^{2}}-\frac{2 z^{2}}{L^{2}}-f(\kappa)
\left(1-\frac{3}{2} \frac{\rho^{2}-2 z^{2}}{R^{2}-L^{2}} \right)
\right)
\end{equation}
where $\ds{f(\kappa)}$ for oblate case, ($\ds{ \kappa>1}$), is
given by
\begin{equation}\label{lnjbhf}
f(\kappa)=\frac{2+\kappa^{2}(4-6~
\frac{\arctan{\sqrt{\kappa^{2}-1}}}{\sqrt{\kappa^{2}-1}}~)}{2
(1-\kappa^{2})}
\end{equation}

\section{Vortex Dipole}

Vortices can be nucleated in a BEC by a localized potential moving
at a velocity above a critical value \cite{hýz6}. Vortices with
opposite circulation are generated at opposite sides of the
condensate. The motion of each vortex arises from its neighbor. In
the presence of weak dissipation, the two vortices slowly drift
together and annihilate when the vortex
separation is comparable with the vortex core size.\\
The critical velocity expression for a vortex dipole formation
derived by Crescimanno \textit{et al.} \cite{enerji6} was found in
good agrement with the experimental observation \cite{ilkdeney5}.
We now extend their approach to include dipolar interaction.
Consider a pair of single vortices with opposite charge. We
suppose the vortices are located symmetrically about the trap
center. Let the distance between the cores be represented by
$\ds{d}$. Substituting $\ds \psi(\textbf{r})=\sqrt{n(\textbf{r})}
e^{i \phi}$ into the GP equation and equating imaginary and real
terms leads to the following hydrodynamics equations:
\begin{equation}\label{juyjyj}
\left(-\nabla^{2}+(\nabla\phi)^{2}+\frac{\rho^{2}+\gamma^{2}
z^{2}}{4}+g' n+\Phi_{dd}\right)\sqrt{n}=\mu^{\prime} \sqrt{n}
\end{equation}
\begin{equation}\label{ýluluu}
-\sqrt{n}~\nabla^{2} \phi-2 \nabla \phi{\cdot}\nabla\sqrt{n}=0
\end{equation}
The term $\ds \nabla^{2} \sqrt{n}$ in the equation (\ref{juyjyj})
is negligible within TF approximation. The ansatz for the phase
function for a vortex dipole is \cite{enerji6}
\begin{equation}\label{mýýouhughu}
\phi(\textbf{r})=l\left( \arctan(~\frac{\sin \theta-\frac{d}{ 2
\rho}}{\cos \theta})- \arctan (\frac{\sin \theta+\frac{d}{2
\rho}}{\cos \theta}~)\right)
\end{equation}
where $ \ds \rho$ and $\ds \theta$ are the polar coordinates, and
$\ds{l}$ is vorticity. The condensate velocity is given by $\ds
v(\textbf{r})=\frac{\hbar}{m}
\nabla \phi (\textbf{r})$.\\
We assume that the vortices are far enough from each other, but
near enough to the trap center. The ansatz is equivalent to
requiring $\ds\frac{1}{\mu^{\prime}} <d^{2}< \mu^{\prime}$
\cite{enerji6}. Using (\ref{mýýouhughu}) we find
\begin{equation}\label{lmnbkg}
|\nabla \phi|=\left(\frac{l^2 d^{2}}{\rho^{2} d^{2}\cos^{2} \theta
+(\rho^{2}-\frac{d^{2}}{4})^{2}+\eta}\right)^{\frac{1}{2}}
\end{equation}
where $\ds\eta=\frac{ l^2 d^{2}}{\mu^{\prime}}$. Excluding the
vortex core regions from the domain complicates the analytic
evaluation of energy and impulse precisely where the TF
approximation fails. To prevent this difficulty, $\ds\eta$ is
added to the denominator of (\ref{lmnbkg}) \cite{enerji6}. This
regulated expression is confirmed by the observation that for
vortex pair not too far from the trap center ($\ds{
d^2<\mu^{\prime} }$), the contribution to
$\ds{\sqrt{n}(\textbf{r})}$ from the kinetic energy term
$\ds{|\nabla \phi|^{2}}$ is never larger than $\ds \mu^{\prime}$ \cite{enerji6}.\\
It is reasonable to approximate the TF density in h.o.u. by
\begin{equation}\label{klmon}
n(\textbf{r})=\frac{1}{g'}\left(\mu'-\frac{\rho^{2}+\kappa^{2}
z^{2}}{4 }\right)
\end{equation}
The correction due to the term $\ds{|\nabla{\phi}|^2}$ is at the
order of $\ds{d^2/R^4}$, which is very small compared to
$\ds{R^2}$ in h.o.u. Let us calculate the total energy and the
total impulse of the condensate. The expression of total energy of
the condensate in h.o.u is given by
\begin{equation}\label{ljöjkklý}
E=\int d^{3}\textbf{r} \left((\nabla \sqrt{n})^{2}+(\sqrt{n}
\nabla\phi)^{2}+\frac{(\rho^{2}+\gamma^{2}z^{2})}{4} n+\frac{g'
}{2}n^{2}+\frac{\Phi_{dd}^{\prime}}{2} n \right)
\end{equation}
The total expression of impulse using the momentum of the
condensate $\ds {\textbf{P}=\frac{i \hbar}{2}[(\nabla\psi^{\ast})
\psi- \psi^{\ast} \nabla \psi]}$ is
\begin{equation}\label{mjnn}
I=\int d^{3}r |\textbf{P}|=\hbar \int d^{3} \textbf{r}
~n(\textbf{r}) ~|\nabla\phi|
\end{equation}
The critical velocity for vortex pair creation using the Landau
criterion is defined as \cite{enerji6}
\begin{equation}\label{kjnjbby}
v_{c}=\frac{E_{l}-E_{0}}{I_{l}}
\end{equation}
Here $\ds {E_{l}}$ and $\ds {I_{l}}$ are the energy and impulse of
vortex state whereas $\ds{ E_{0}}$ is the energy of non-vortex
state. Analytical evaluation of these energy, impulse and critical
velocity functions were performed by Crescimanno \textit{et al.}
in two dimensions for a nondipolar condensate \cite{enerji6}. In
this study, we will calculate the critical velocity of a dipolar
condensate in an oblate trap. As the resulting equations for $\ds
v_{c}$ are complicated, it is necessary to obtain them numerically
for given $\ds a_{s} $, $\ds \varepsilon_{dd}$, $\ds \kappa$. We
compare the critical velocities of dipolar and non-dipolar
condensates.

\section{Results}

\begin{figure}
\centering
\includegraphics[width=6cm]{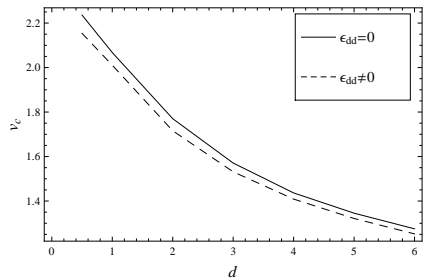}
\caption{The critical velocity, $\ds{v_c}$ ($mm/s$), for
$\ds{\varepsilon_{dd}=0}$ (solid) and $\ds{\varepsilon_{dd}=0.15}$
(dashed) as a function of $\ds{d}$ in units of h.o.u. for an
oblate trap with $\ds{\gamma=5}$ } \label{uzaklik}
\end{figure}\begin{figure}
\centering
\includegraphics[width=6cm]{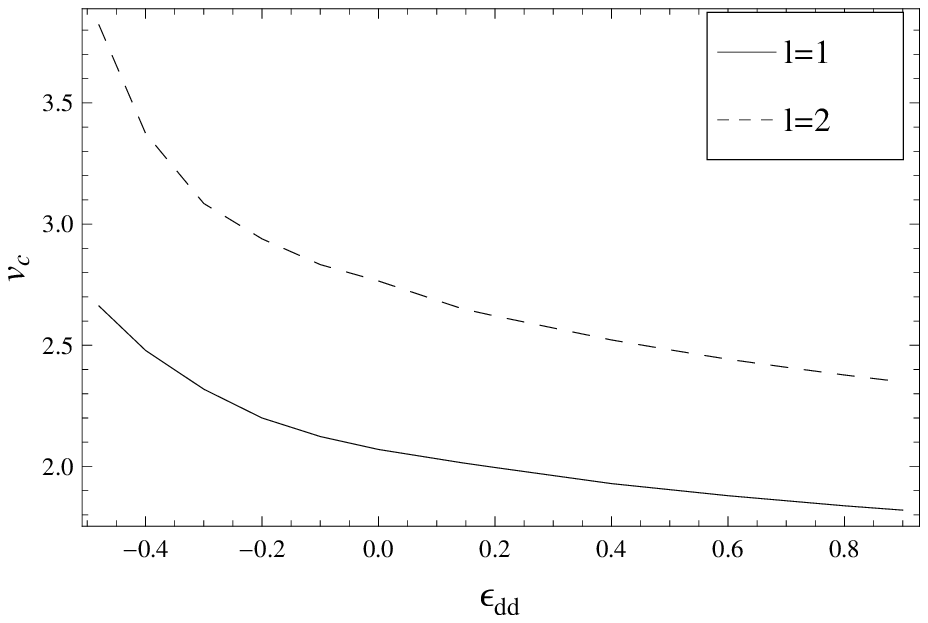}
\caption{The critical velocity, $\ds{v_c}$ ($mm/s$), for
$\ds{d=1}$ in units of h.o.u. as a function of
$\ds{\varepsilon_{dd}}$ for an oblate trap with $\ds{\gamma=5}$.
The solid (dashed) curve is for singly (doubly) quantized vortex
dipole. } \label{epsilon}
\end{figure}\begin{figure}
\centering
\includegraphics[width=6cm]{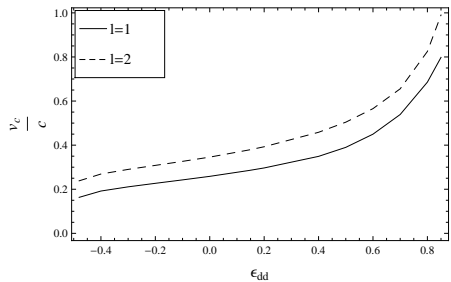}
\caption{The ratio of critical velocity to the speed of sound,
$\ds{\frac{v_c}{c}}$, for $\ds{d=1}$ in units of h.o.u. as a
function of $\ds{\varepsilon_{dd}}$ for an oblate trap with
$\ds{\gamma=5}$. The solid (dashed) curve is for singly (doubly)
quantized vortex dipole.  } \label{epsilon2}
\end{figure}
In this paper, within the TF regime we perform a numerical
calculation of critical velocity for a vortex pair formation in a
dipolar BEC. We examine dipolar gas containing $\ds 150 000$
$\ds{^{52}Cr}$ atoms in an oblate trap with trap frequencies $\ds{
\omega_{\bot}=2 \pi \times 200}$  $rad/s$, $\ds{ \omega_{z}=2 \pi
\times 1000} $ $rad/s$, so the trap aspect ratio is $\ds
\gamma=5$. The magnitude of the scattering length for
$\ds{^{52}Cr}$ is $\ds {105a_{B}}$ ($a_{B}$ is the Bohr magneton).
We assume that vortex separation $\ds{d}$ satisfies the condition
$\ds{\frac{1}{\mu^{\prime}} <d^{2}< \mu^{\prime}}$. The chemical
potential is approximately $42$ in h.o.u. for
$\ds{\varepsilon_{dd}=0}$ and changes very slightly with $\ds{d}$.
The dipole-dipole interaction decreases the chemical potential.
Below, we perform numerical integration of energy and impulse
functional (\ref{ljöjkklý},\ref{mjnn}) to find
the critical velocity of vortex pair formation (\ref{kjnjbby}).\\
It is well known that  in an oblate trap, condensate aspect ratio,
$\ds \kappa$, decreases with increasing $\ds{ \varepsilon_{dd}}$
and the dipole-dipole interaction energy is positive. Minimizing
the energy functional of the dipolar condensate, we find that the
condensate aspect ratio increases slightly with $\ds{d}$ and the
radius $\ds{R}$ is almost the same for all $\ds{d}$. For example,
$\ds{ \kappa}$ is between $\ds{4.73}$ and $\ds{4.76}$ when
$\ds{d}$ is between $\ds{0.5}$ and $\ds{6}$, respectively for
$\ds{\varepsilon_{dd}=0.15}$. The kinetic energy changes
significantly with vortices separation $\ds{d}$, while the other
terms in the energy expression change slightly with $\ds{d}$. The
impulse of vortex state, ${\ds I}$, depends strongly on $\ds{d}$
and $\ds{\varepsilon_{dd}}$ since it scales as $\ds{d/R^2}$
(\ref{lmnbkg},\ref{mjnn}). It decreases with $\varepsilon_{dd}$,
since the dipolar interaction stretches the
cloud radially in an oblate trap.\\
The critical velocity $\ds v_{c}$ decreases with increasing
separation $\ds{d}$ for a non-dipolar condensate \cite{enerji6}.
This is because the TF density is a maximum at the trap center and
reduces with the distance away from the trap center. We expect
that the critical velocity decreases with $\ds{d}$ also for a
dipolar BEC since parabolic form of density retains in the case of
dipolar interaction. Fig-\ref{uzaklik} plots the critical velocity
$\ds v_{c}$ as a function of the distance between the vortices for
$\ds{\varepsilon_{dd}=0}$ (solid curve) and for
$\ds\varepsilon_{dd}=0.15$ (dashed curve). The critical velocity
is between $\ds{ 1.25-2.15} $ $mm/s$ for
$\ds\varepsilon_{dd}=0.15$, and $\ds 1.27-2.24$ $mm/s$ for
$\ds\varepsilon_{dd}=0$ in the range $\ds{6>d>0.5}$. The critical
velocity difference between dipolar and non-dipolar condensates
becomes smaller as $\ds{d}$ is increased. This is because
$\ds{\kappa}$ increases with $\ds{d}$. As can also be seen from
the figure, the inclusion of dipolar interaction decreases the
critical velocity for a fixed value of $\ds{d}$. This is always
true for positive values of $\ds\varepsilon_{dd}$. In the case of
negative values of $\ds\varepsilon_{dd}$, the effect of
dipole-dipole interaction is the increased critical velocity.
Fig-\ref{epsilon} shows the critical velocity as a function of
$\ds\varepsilon_{dd}$ for fixed $\ds{d=1}$. The solid curve shows
singly quantized vortex dipole while the dashed curve shows doubly
quantized vortex dipole. The effect that $\ds{v_c}$ is decreased
with increasing $\ds\varepsilon_{dd}$ in an oblate trap is the
first main result of this paper. It is energetically less
expensive to nucleate a vortex in an oblate dipolar Bose-Einstein
condensate than in a condensate with only contact interactions. At
first sight, this might seem counterintuitive since the
dipole-dipole interaction energy is positive in an oblate trap. It
is remarkable to note that although dipolar interaction is
positive for an
oblate trap, the excess dipolar energy is negative.\\
The nucleation of multiply charged vortex dipoles was observed for
trap translation velocities well above $\ds{v_c}$ in the
experiment \cite{ilkdeney5}. Furthermore, it was observed that the
vortices exhibit periodic orbital motion and vortex dipoles may
exhibit lifetimes of many seconds, much longer than a single
orbital period  \cite{ilkdeney5}. Fig-\ref{epsilon} compares the
critical velocity for singly and doubly quantized vortex dipole.
As expected, $\ds{v_c}$ is bigger
for doubly quantized vortices.\\
On the investigation of a vortex dipole, not only $\ds{v_c}$, but
also the ratio $\ds{v_c/c}$ is of importance. Here $\ds{c}$ is the
speed of sound \cite{rev2}
\begin{equation}\label{kjng}
c=\sqrt{\frac{n_0}{m} g\left(1+\epsilon_{dd}(3
\cos^2{\alpha}-1)\right)}
\end{equation}
where $\ds{\alpha}$ is the angle between directions of the wave
vector and the dipoles. Suppose the direction of the phonon wave
vector is perpendicular to the orientation of the dipoles
($\ds{\alpha=\pi/2}$). In this case, the speed of sound becomes
$\ds{c=\sqrt{\frac{n_0}{m} g(1-\epsilon_{dd})}}$. Remarkably, both
$\ds{v_c}$ and $\ds{c}$ decrease with $\ds{\varepsilon_{dd}}$.
However, the ratio $\ds{v_c/c}$ increases with
$\ds{\varepsilon_{dd}}$. As $\ds{\varepsilon_{dd}}$ goes to one,
the critical velocity approaches to the speed of sound. In
Fig-\ref{epsilon2}, we plot $\ds{v_c/c}$ versus $\ds{
\varepsilon_{dd}}$ for singly (solid curve) and doubly quantized
vortex dipoles (dashed curves) for fixed $\ds d=1$. The effect
that $\ds{v_c/c}$ is increased with increasing
$\ds\varepsilon_{dd}$ in an oblate trap is the second main result
of this paper. The ratio $\ds{v_c/c}$ for singly quantized
vortices are found to be between $\ds 0.16-0.31$ for dipolar
condensate with $\ds{\varepsilon_{dd}=0.15}$ and $\ds 0.15-0.28$
for non-dipolar condensate in the range $\ds{6>d>0.5}$. As
expected, the ratio $\ds{v_c/c}$ increases for doubly quantized
vortices.\\
In this paper, we have studied single and multiply quantized
vortex dipoles in an oblate dipolar Bose Einstein condensate. We
have shown that $\ds{v_c}$ is decreased while $\ds{v_c/c}$ is
increased with increasing $\ds\varepsilon_{dd}$  in an oblate
trap. The dynamics of vortex dipole in a dipolar BEC is worth
studying. Helpful discussions with A. Kilic are gratefully
acknowledged.


\begin{thebibliography}{99}
\bibitem{jack} B. Jackson, J. F. McCann, and C. S. Adams, Phys. Rev. Lett. {\bf 80}, 3903 (1998).
\bibitem{int13} C. Raman, M. Kohl, R. Onofrio, D. S. Durfee, C. E. Kuklewicz, Z. Hadzibabic, and W. Ketterle, Phys. Rev. Lett. {\bf83}, 2502 (1999).
\bibitem{int40} R. Onofrio, C. Raman, J. M. Vogels, J. Abo-Shaeer, A. P.Chikkatur, and W. Ketterle, Phys. Rev. Lett. {\bf85},2228,(2000) .
\bibitem{63} B. Jackson, J. F. McCann, and C. S. Adams, Phys. Rev. A {\bf61}, 051603(R) (2000).
\bibitem{ilkdeney5} T. W. Neely, E. C. Samson, A. S. Bradley, M. J. Davis, and B. P. Anderson,  Phys. Rev. Lett. {\bf104}, 160401(2010).
\bibitem{enerji6} M. Crescimanno, C. G. Koay, R. Peterson, R. Walworth Phys. Rev. A {\bf62}, 063612 (2000).
\bibitem{int5} L. Landau, J. Phys. (Moscow)  {\bf58}, 71 (1941).
\bibitem{int24} C. Raman, J. R. Abo-Shaeer, J. M. Vogels, K. Xu, and W. Ketterle,  Phys. Rev. Lett. {\bf87}, 210402 (2001).
\bibitem{int41} J. S. Stiebberger, and W. Zwerger, Phys. Rev. A {\bf62}, 061601, (2000).
\bibitem{int14} N. Pavloff, Phys. Rev. A  {\bf66}, 013610 (2002).
\bibitem{fre} D. V. Freilich, D. M. Bianchi, A. M. Kaufman, T. K. Langin, and D. S. Hall, Science {\bf6329}, 1182 (2010).
\bibitem{fre2} J. A. Seman, et al. Phys. Rev. A {\bf 82}, 033616 (2010).
\bibitem{int15} J. P. Martikainen, K. A. Suominen, L. Santos, T. Schulte, and A. Sanpera, Phys. Rev. A  {\bf64}, 063602 (2001).
\bibitem{int16} M. Liu, l. H. Wen, H. W. Xiong, and M. S. Zhan, Phys. Rev. A  {\bf73}, 063620 (2006).
\bibitem{int17} R. Geurts, M. V. Milosevic, and F. M. Peeters, Phys. Rev. A  {\bf78}, 053610 (2008).
\bibitem{int18} D. Schumayer, D. A. W. Hutchinson, Phys. Rev. A  {\bf75}, 015601 (2007).
\bibitem{pekko} P. Kuopanportti, Jukka A. M. Huhtamaki, and M. Mottonen, arXiv:1011.1661.
\bibitem{int19} L. C. Crasovan, V. Vekslerchik, V. M. Perez-Garcýa, J. P. Torres, D. Mihalache, and L. Torner1, D. A. W. Hutchinson, Phys. Rev. A  {\bf68}, 063609 (2003).
\bibitem{int20} L. D. Carr, and C. W. Clark, Phys. Rev. Lett.  {\bf97}, 010403 (2006).
\bibitem{int21} V. Pietilä, V. Pietilä, T. Isoshima, J. A. M. Huhtamäki, and S. M. M. Virtanen, Phys. Rev. Lett.  {\bf74}, 023603 (2006).
\bibitem{int22} W. Li, M. Haque, and S. Komineas, Phys. Rev. A  {\bf77}, 053610 (2008).
\bibitem{int23} Q. Zhou, and H. Zhai, Phys. Rev. A  {\bf70}, 043619 (2004).
\bibitem{deney01} A. Griesmaier, J. Werner, S. Hensler, J. Stuhler, and T. Pfau, Phys. Rev. Lett. {\bf 94}, 160401 (2005).
\bibitem{deney02} T. Lahaye, T. Koch, B. Frohlich, M. Fattori, J. Metz, A. Griesmaier, S. Giovanazzi, and T. Pfau, Nature (London) {\bf 448}, 672 (2007).
\bibitem{deney03} T. Koch, T. Lahaye, J. Metz, B. Frohlich, A. Griesmaier, and T. Pfau, Nat. Phys. {\bf 4}, 218 (2008).
\bibitem{rev1} M.A. Baranov, Phys. Rep. {\bf464}, 71  (2008).
\bibitem{rev2} T Lahaye, C Menotti, L Santos, M Lewenstein, T Pfau, Rep. Prog. Phys. {\bf 72}, 126401 (2009).
\bibitem{enerji1} K. Goral, K. Rzazewski, and T. Pfau, Phys. Rev. A  {\bf61}, 051601(R) (2000).
\bibitem{enerji2} S. Yi and L. You, Phys. Rev. A {\bf61}, 041604(R) (2000).
\bibitem{int35} R. M. W. van Bijnen, A. J. Dow, D. H. J. O'Dell, N. G. Parker, and A. M. Martin,  Phys. Rev. A {\bf80}, 033617 (2009).
\bibitem{int37} R. M. W. van Bijnen, N. G. Parker, S. J. J. M. F. Kokkelmans, A. M. Martin, and D. H. J. O'Dell, Phys. Rev. A {\bf82}, 033612 (2010).
\bibitem{int39} S. Giovanazzi, A. Go¨rlitz, and T. Pfau, Phys. Rev. Lett. {\bf89}, 130401 (2002).
\bibitem{int36} D. H. J. O'Dell, S. Giovanazzi, and C. Eberlein, Phys. Rev. Lett. {\bf92},250401  (2004).
\bibitem{int29} R. M. W. van Bijnen, D. H. J. O'Dell, N. G. Parker, and A. M. Martin , Phys. Rev. Lett. {\bf98}, 150401 (2007).
\bibitem{int30}  S. Yi ,and H. Pu, Phys. Rev. A {\bf73}, 061602(R)  (2006).
\bibitem{cem} C. Yuce, arXiv:1012.5392.
\bibitem{kal0} M. Klawunn, R. Nath, P. Pedri, and L. Santos, Phys. Rev. Lett. {\bf100}, 240403 (2008).
\bibitem{kal} M Klawunn and L Santos, New J. Phys. {\bf711} 055012 (2009).
\bibitem{abad0} M. Abad, M. Guilleumas, R. Mayol, M. Pi, D. M. Jezek, Phys. Rev. A {\bf81}, 043619 (2010).
\bibitem{abad1} M. Abad, M. Guilleumas, R. Mayol, M. Pi and D. M. Jezek, Laser Physics {\bf 20} 1190 (2010).
\bibitem{int28} M. Abad., M. Guilleumas, R. Mayol, and M. Pi, Phys. Rev. A {\bf79}, 063622  (2009).
\bibitem{int31}  D. H. J. O'Dell, and C. Eberlein,Phys. Rev. A,  {\bf75}, 013604 (2007).
\bibitem{int32}  C. Yuce, and Z. Oztas, J. Phys. B: At. Mol. Opt. Phys. {\bf43}  (2010).
\bibitem{int33}  R. Y. Wilson , S. Ronen, and J. L. Bohn, Phys. Rev. A {\bf79}, 013621 (2009).
\bibitem{int34} T. Isoshima, J. Huhtamäki, and M. M. Salomaa,  Phys. Rev. A {\bf68}, 033611 (2003).
\bibitem{int38} R. M. Wilson, S. Ronen, and J. L. Bohn,  Phys. Rev. Lett. {\bf104}, 094501 (2010).
\bibitem{int25} C. Eberlein, S. Giovanazzi, and  D. H. J. O'Dell , Phys. Rev. A {\bf71}, 033618  (2005).
\bibitem{hýz6} N. G. Parker, B. Jackson, A. M. Martin, and C. S. Adams, Springer Series on Atomic, Optical, and Plasma Physics  {\bf45}, VI, 173-189 (2008).
\end{thebibliography}
\end{document}